\documentclass[12pt,preprint]{aastex}

\shorttitle{HSA Observations of the $z=1.87$ SMG GOODS~850--3}
\shortauthors{Momjian et al.}

\begin{document}

\title{High Sensitivity Array Observations of the $z=1.87$ Sub-Millimeter Galaxy GOODS~850--3
\footnote{Based in part on observations obtained at the Canada-France-Hawaii Telescope
(CFHT), which is operated by the National Research Council of Canada, the
Institut National des Sciences de l'Univers of the Centre National de la Recherche
Scientifique of France, and the University of Hawaii, and the Subaru Telescope,
which is operated by the National Astronomical Observatory of Japan.
}}

\author{Emmanuel Momjian}
\affil{National Radio Astronomy Observatory, P. O. Box O, Socorro, NM, 87801, USA}
\email{emomjian@nrao.edu}

\author{Wei-Hao Wang\altaffilmark{a}}
\affil{Academia Sinica Institute of Astronomy and Astrophysics, P.O. Box 23-141, Taipei 10617, Taiwan}
\email{whwang@asiaa.sinica.edu.tw}
\altaffiltext{a}{National Radio Astronomy Observatory, P. O. Box O, Socorro, NM, 87801, USA}

\author{Kirsten K. Knudsen}
\affil{Argelander-Institut f\"{u}r Astronomie, Auf dem H\"{u}gel 71, D-53121 Bonn, Germany}
\email{knudsen@astro.uni-bonn.de}

\author{Christopher L. Carilli}
\affil{National Radio Astronomy Observatory, P. O. Box O, Socorro, NM, 87801, USA}
\email{ccarilli@nrao.edu}

\author{Lennox L. Cowie}
\affil{Institute for Astronomy, University of Hawaii, 2680 Woodlawn Drive, Honolulu, HI 96822, USA}
\email{cowie@ifa.hawaii.edu}

\author{Amy J. Barger\altaffilmark{b,c}}
\affil{Department of Astronomy, University of Wisconsin-Madison, 475 North Charter St., Madison, WI 53706}
\email{barger@astro.wisc.edu}
\altaffiltext{b}{Department of Physics and Astronomy, University of Hawaii}
\altaffiltext{c}{Institute for Astronomy, University of Hawaii}

\begin{abstract}
We present sensitive phase-referenced VLBI results on the radio continuum emission from
the $z=1.87$ luminous submillimeter galaxy (SMG) GOODS~850--3. The observations were
carried out at 1.4~GHz using the High Sensitivity Array (HSA).
Our sensitive tapered VLBI image of GOODS~850--3 at 0.47$''$ $\times$ 0.34$''$ ($3.9 \times 2.9$~kpc) resolution
shows a marginally resolved continuum structure with a peak
flux density of $148 \pm 38~\mu$Jy~beam$^{-1}$, and a total flux density of $168 \pm 73~\mu$Jy,
consistent with previous VLA and MERLIN measurements. 
The deconvolved size of the source is $0\rlap{.}{''}27 (\pm 0\rlap{.}{''}12) \times < 0\rlap{.}{''}23$, or
$2.3 (\pm 1.0) \times < 1.9$~kpc${^2}$,
and the derived intrinsic brightness temperature is $> (5 \pm 2) \times 10^3$~K. 
The radio continuum position of this galaxy coincides with a bright and extended
near-infrared source that nearly disappears in the deep HST optical image, indicating a dusty
source of nearly 9~kpc in diameter.
No continuum emission is detected at the full VLBI resolution ($13.2 \times 7.2$~mas, $111 \times 61$~pc),
with a $4\sigma$ point source upper limit of 26~$\mu$Jy~beam$^{-1}$, or an upper limit to the
intrinsic brightness temperature of $4.7 \times 10^5$~ K.
The extent of the observed continuum source at 1.4~GHz and the derived
brightness temperature limits are consistent with the radio emission (and thus presumably the far-infrared
emission) being powered by a major starburst in GOODS~850--3, with a star formation
rate of $\sim$2500~$M_{\odot}~{\rm yr}^{-1}$. 
Moreover, the absence of any continuum emission at the full resolution of the VLBI observations indicates the
lack of a compact radio AGN source in this $z=1.87$ SMG.

\end{abstract}


\section{INTRODUCTION}

The strength of the far-infrared (FIR) extragalactic background implies that
a large fraction of galaxy formation and black hole accretion
activity is hidden by dust (see a review in \citealp{lagache05}).
A large fraction ($\sim30\%$) of the background has been resolved in
the submillimeter into discrete high-redshift sources
\citep[e.g.,][]{smail97,hughes98,barger98}.  Bright ($\gtrsim5$ mJy
at 850 $\mu$m) submillimeter galaxies (SMGs) were found to be mostly
at high redshifts of $z\sim1.5$--3, and they dominate the total star
formation in this redshift range \citep{chap03,chap05}.
They have total infrared (IR) luminosities of $10^{12}$ to $>10^{13}$
$L_\sun$, corresponding to star formation rates of $10^2$-$10^3$ $M_\sun$ yr$^{-1}$.

An outstanding question about bright SMGs is whether the FIR
emission is powered by active galactic nuclei (AGNs) or by starbursts.
This has been studied at various wavelengths using X-ray
\citep{alexander03,alexander05}, optical \citep{barger99,chap05},
near-infrared (NIR) \citep{swinbank04}, and mid-infrared (MIR) data \citep{pope08,murphy09}.
At even longer wavelengths, radio and mm interferometers add another powerful way
to investigate the existence of AGNs through morphology.
A small sample of two SMGs were marginally resolved by the Submillimeter Array
at sub-arcsec resolutions, suggesting extended starbursts to be the main
power source \citep{younger}.  By combining the Multi-Element
Radio Linked Interferometer (MERLIN) and the Very Large Array (VLA),
\citet{chap04} obtained high angular resolution ($\sim0\farcs3$)
1.4 GHz images of 12 SMGs.  The majority (8 out of 12; 67\%) of their sources were
resolved by MERLIN+VLA, again suggesting extended starbursts.  However, the
resolution of MERLIN+VLA is insufficient to probe radio emission from compact
radio AGNs. Whether the remaining 33\% of compact radio sources in the
MERLIN+VLA sample are powered by compact nuclear starbursts or by AGNs
thus remains to be tested.

The high resolution of Very Long Baseline Interferometry (VLBI) observations permits
a detailed look at the physical structures in the most distant cosmic sources. Also,
sensitive VLBI continuum observations can be used to determine the nature of the energy
source(s) in these galaxies at radio frequencies. To date, several high redshift QSOs
have been imaged at milliarcsecond
resolution \citep{FRE97,FRE03,BEE04,MOM04,MOM05,MOM08}. These are the highest resolution
studies of such distant QSOs by far. Here we are expanding such work on high-$z$
SMGs. In this paper, we present sensitive VLBI
observations of the SMG GOODS~850--3.

The source GOODS~850--3 (aka SMM~J123618.3+621551, GN6, \citealp{chap04,pope06}),
which has an optical redshift of $1.865$\footnote{\citealp{pope08} derived a redshift
of $2.00\pm0.03$ from \emph{Spitzer} MIR spectroscopy with PAH features. 
Our observations in \S~4.2 show that GOODS~850--3 is extremely optically faint
and therefore the spectroscopic observations of \citet{chap05} might have been made
on another galaxy.  Nevertheless, the inconsistency between the
two redshift measurements does not affect our analysis.}  \citep{chap05},
is a luminous SMG with $S_{\rm 850\mu m}=7.72 \pm 1.02$~mJy \citep{WCB04}.
This object is one of the strongest radio sources among SMGs, and it is
designated as a compact source at $0.3''$ resolution \citep{chap04,chap05}.
The flux density of GOODS~850--3 at 1.4~GHz, measured with MERLIN+VLA observations, is
$S_{\rm 1.4 GHz} = 151 \pm 11~\mu$Jy.

Throughout this paper, we assume a flat cosmological model with
$\Omega_{m}=0.3$, $\Omega_\Lambda=0.7$, and
${H_{0}=70}$~km~s$^{-1}$~Mpc$^{-1}$. In this model, and at the distance of GOODS~850--3,
1~milliarcsecond (mas) corresponds to 8.4~pc.

\section{OBSERVATIONS AND DATA REDUCTION}

The VLBI observations of GOODS~850--3 were carried out at 1.4~GHz on 2006 December 9 and
2007 February 13 using the High Sensitivity Array (HSA), which includes
the Very Long Baseline Array (VLBA), the phased Very Large Array (VLA), and the
Green Bank Telescope (GBT) of the NRAO\footnote{The National Radio Astronomy
Observatory is a facility of the National Science Foundation operated under
cooperative agreement by Associated Universities, Inc.}. Eight adjacent 8~MHz
baseband channel pairs were used in the observations, both with right- and
left-hand circular polarizations, and sampled at two bits. The data were
correlated at the VLBA correlator in Socorro, NM, with a 2~s correlator
integration time. The total observing time was 16~hr. 
In these observations, the shortest baseline is between the phased VLA and the VLBA antenna in
Pie Town, NM, which is 52~km. This short-spacing limit filters out all spatial structure
larger than about $0\rlap{.}{''}42$. Table~1 summarizes the parameters of these observations.

The observations employed nodding-style phase referencing using the calibrator J1229+6335
($S_{\rm 1.4 GHz} = 0.38$~Jy) with a cycle time of 280~sec, 200~sec on the target source and
80~sec on the calibrator. The angular separation between the target and the phase calibrator
is $1.55\degr$. A number of test cycles were also included to monitor the coherence of the phase
referencing. These tests involved switching between two calibrators, the phase calibrator
J1229+6335 and the phase-check calibrator J1219+6600 ($S_{\rm 1.4 GHz} = 0.06$~Jy), using a
similar cycle time to that used for the target source. The angular separation between these two
calibrators is $2.6\degr$.

The accuracy of the phase calibrator position is important in phase-referencing
observations \citep {WAL99}, as this determines the accuracy of the absolute
position of the target source and any associated components. Phase referencing,
as used here, is known to preserve absolute astrometric positions to better
than $\pm 0\rlap{.}{''}01$ \citep{FOM99}.

Data reduction and analysis were performed using the Astronomical Image
Processing System (AIPS) of the NRAO.  After applying {\it a priori} flagging, amplitude
calibration was performed using measurements of the antenna gain and system
temperature for each station. Ionospheric corrections were applied using the
AIPS task ``TECOR''. The phase calibrator J1229+6335 was self-calibrated in
both phase and amplitude and imaged in an iterative cycle.

Images of the phase-check calibrator, J1219+6600, were deconvolved using two
different approaches: (a) by applying the phase and the amplitude
self-calibration solutions of the phase reference source J1229+6335
(Figure~1{\it{a}}), and (b) by self calibrating J1219+6600 in both
phase and amplitude (Figure~1{\it{b}}). The peak surface brightness ratio of
the final images from the two approaches gives a measure of the effect of
residual phase errors after phase referencing (i.e., `the coherence' due to
phase referencing). At all times, the coherence was found to be better than
93\%.

The self-calibration solutions of the phase calibrator, J1229+6335, were applied on the
target source, GOODS 850--3, which was then deconvolved and imaged at various spatial
resolutions by tapering the visibility data.

\section{RESULTS \& ANALYSIS}

Imaging the target source at the full resolution of the VLBI array, which is $13.2 \times 7.2$~mas
($111 \times 61$~pc, PA=$-16\degr$), achieved an rms noise level of $6.5~\mu$Jy~beam$^{-1}$,
but it did not reveal
any continuum component in the field of GOODS~850--3. This indicates the absence of any
compact radio continuum emission with flux densities of $\geq 4\sigma$ or $26~\mu$Jy~beam$^{-1}$,
which in turn implies an upper limit to the intrinsic brightness temperature (corresponding to a rest
frequency of $\sim4$~GHz) of $4.7 \times 10^5$~K for any compact radio source in GOODS~850--3. Our
coherence tests during these observations using two VLBI calibrators show that the lack of
a strong point source in GOODS~850--3 at the full resolution of the array cannot be due to
the phase referencing procedure. The VLBI flux limit reported above is almost an order of
magnitude lower than the flux measured by the VLA+MERLIN ($151 \pm 11~\mu$Jy) at 1.4 GHz
\citep{chap04}. This immediately implies that more than 90\% of the radio continuum emission in
GOODS~850--3 is extended and not confined to a central AGN.

In the following, we assess whether our HSA observations can recover the flux seen by the
VLA and MERLIN. To do so, we applied two dimensional Gaussian tapers on the visibility
data with various values. However, the only image with a reliable $\sim 4\sigma$, detection was
obtained by applying a Gaussian taper falling to 30\% at 0.5 M$\lambda$ in both the u- and v-
directions. This gave a beam size of $0\rlap{.}{''}47 \times 0\rlap{.}{''}34$ ($3.9 \times 2.9$~kpc${^2}$, P.A.=$18\degr$). The resulting image is shown in Figure~2. This image is practically made from the shortest
baseline in our data set (phased VLA $-$ Pie Town), but only after calibrating its visibilities using all
the antennas in the array as described in \S 2. The rms noise level in this naturally weighted image
is 38~$\mu$Jy~beam$^{-1}$. At this resolution, a Gaussian model fitting reveals a marginally resolved
continuum source
with a peak flux density of $148 \pm 38~\mu$Jy~beam$^{-1}$ and a total flux density of $168 \pm 73~\mu$Jy,
which agrees well with the flux density measured with the VLA+MERLIN at 1.4 GHz \citep{chap04}.
Deconvolving the synthesized beam from the Gaussian fitting model of the source results in
$0\rlap{.}{''}27 \pm 0\rlap{.}{''}12$ for the size of the major axis,
and an upper limit of $0\rlap{.}{''}23$ for the size of the minor axis, i.e.,
$0\rlap{.}{''}27 (\pm 0\rlap{.}{''}12) \times < 0\rlap{.}{''}23$, or
$2.3 (\pm 1.0) \times < 1.9$~kpc${^2}$. The derived intrinsic brightness temperature limit is
$T_{\rm b} > (5 \pm 2) \times 10^3$~K.

\section{DISCUSSION}

\subsection {Radio Properties}

We have detected 1.4~GHz emission from GOODS~850--3 (corresponding to a rest-frame 
frequency of $\sim$4~GHz) using the HSA. At a relatively low spatial resolution
($0\rlap{.}{''}47 \times 0\rlap{.}{''}34$; Figure~2)
the limit on the intrinsic brightness temperature value of the detected continuum source is
$> (5 \pm 2) \times 10^3$~K. Its flux density at this resolution
is consistent with that measured with the VLA+MERLIN \citep{chap04,chap05}.
This implies that the radio continuum emission at
1.4~GHz is confined to the extent of the structure seen in the VLA+MERLIN
image at $0\rlap{.}{''}3$ resolution \citep{chap04}, or to the extent of the deconvolved
size scale reported in this paper,
which is $0\rlap{.}{''}27 (\pm 0\rlap{.}{''}12) \times < 0\rlap{.}{''}23$, or
$2.3 (\pm 1.0) \times < 1.9$~kpc${^2}$.

At the full resolution of our array ($13.2 \times 7.2$~mas), the radio emission from
GOODS~850--3 is resolved out and does not show any single dominant source of
very high brightness temperature ($< 4.7\times 10^5$ K). 
This is in contrast to
the results obtained by \citet{MOM04} on a sample of three high-$z$ radio-loud
quasars imaged with the VLBA, namely J1053-0016 ($z=4.29$), J1235-0003 ($z=4.69$),
and J0913+5919 ($z=5.11$). In each of these $z>4$ quasars, a
radio-loud AGN dominates the emission at 1.4~GHz on a few mas size scale, with
intrinsic brightness temperatures in excess of $10^9$~K. 

\citet{CON91} have derived an empirical upper limit to the brightness
temperature for nuclear starbursts. For a frequency value of 4~GHz (our rest frequency),
the resulting limit on the intrinsic brightness temperature is $T_{\rm b} \leq 4.8 \times 10^4$~K,
while typical radio-loud AGNs have brightness temperatures exceeding this value by at least
two orders of magnitude. These authors also present a possible physical model
for this limit involving a mixed non-thermal and thermal radio emitting (and
absorbing) plasma, constrained by the radio-to-FIR correlation for star-forming
galaxies. The measured intrinsic brightness temperature limit for the radio source in GOODS~850--3
is consistent with the empirical upper limit value for nuclear starbursts.

For GOODS~850--3, the derived intrinsic brightness
temperatures from our VLBI observations are typical of starburst galaxies.
However, when compared to local starburst powered Ultra-Luminous IR galaxies
(ULIRGs; Sanders \& Mirabel 1996), such as Arp~220, Mrk~273, and
IRAS~17208--0014 \citep{SMI98,CAR00,MOM03,MOM06,LO06}, the radio continuum emission from
GOODS~850--3 is an order of magnitude greater in luminosity, but also 
an order of magnitude larger in extent.

In starburst dominated galaxies, the radio continuum is the sum of
supernovae (SNe), supernova remnants, and residual relativistic electrons in the interstellar
medium, as shown in the model presented by \citet{CON91}. However, detecting
individual SNe in GOODS~850--3 at $z =1.87$ is unlikely.
\citet{LO06} reported the detection of 49 luminous radio SNe in the
prototype ULIRG Arp~220 with flux densities between 0.053 and 1.228 mJy and
typical upper limits on their linear extent of less then 1~pc. At the distance of GOODS~850--3, the
flux densities of such luminous radio SNe would be between $ (5-115) \times
10^{-3}~\mu$Jy. These values are two to three orders of magnitude lower that the
rms noise levels achieved in our VLBI observations.

Our radio continuum results with the HSA clearly rule out a compact radio-loud AGN in GOODS~850--3
and show physical characteristics consistent with an extreme starburst. However, 
we cannot completely rule out that the radio emission in this source is from a kpc scale radio-jet structure
with a very faint compact AGN component that contributes 
to less than 10\% of the total radio continuum emission, and hence falls below our
detection threshold at the full resolution of the array.
Therefore, we further extend the discussion regarding the nature of the power source(s) in this SMG,
and whether it hosts a radio-quiet AGN, by looking into the optical and IR properties
of GOODS~850--3 at various bands, as described in the following section.

\subsection {Multi-wavelength Properties}

Evidence for an intensive starburst in GOODS~850--3 is also seen at various other wavelengths.
First, \citet{carilli99} presented a method of using the local radio--FIR correlation
and the radio and submillimeter flux densities to estimate redshifts of SMGs.
\citet{barger00} derived a simple redshift formula for
Arp 220-like Spectral Energy Distribution (SEDs): $z+1=0.98(S_{850}/S_{1.4})^{0.26}$. With this formula,
the measured redshift $z=1.865$, and the radio and submillimeter fluxes of GOODS 850--3, we find
its radio and submillimeter SED to be consistent with Arp 220 within 20\%.
This implies that this SMG excellently follows the same radio--FIR correlation as Arp 220, and
suggests that its FIR emission is powered by a starburst.

GOODS~850--3 is among the first SMGs with an X-ray detection \citep{alexander03}.
It is detected at soft X-rays in the 2Ms \emph{Chandra} image of the Chandra Deep Field-North (CDFN),
but it is not detected at hard X-rays. It has an X-ray luminosity of $\sim0.5\times10^{42}$
erg s$^{-1}$. These authors suggested that this source is powered by star formation, but
they did not rule out a highly obscured AGN with low X-ray luminosity.

\citet{pope08} obtained a MIR spectrum on GOODS 850--3 (GN06 in their paper)
and decomposed the spectrum into PAH and ``continuum'' components.
The continuum component contributes $\sim50\%$ to its MIR luminosity,
and such a component can be powered by a dusty AGN. The MIR spectral
decomposition method was further refined by \citet{murphy09}, who noted
that such MIR continuum fraction is a strict upper limit on the AGN contribution
at this wavelength band.
However, AGN dust components are warmer than those powered by starbursts and
contribute mainly to the MIR.  We therefore expect the AGN fraction in the
total IR luminosity of GOODS 850--3 to be negligible.

None of the above observational evidence, including our HSA observations, absolutely
rule out the existence of an obscured AGN in GOODS 850--3.  However, a kpc-scale
starburst is consistent with all observations from the X-ray to the radio.

The last piece of evidence comes from the optical and NIR morphologies.
For the optical, we combined the latest version of the four (F435W, F606W, F775W, and F850LP) \emph{HST} 
ACS images from The Great Observatories Origins Deep Survey (GOODS; \citealp{giavalisco04}) to form a deep
``white'' optical image for GOODS 850--3. For the NIR, we used three sets of data.
First, Lihwai Lin et al.\ (in preparation) obtained nearly 30 hr of GOODS-N imaging data at
$J$-band with the Wide-Field Infrared Camera (WIRCam) on the 3.6~m Canada-France-Hawaii Telescope (CFHT).
We downloaded these $J$-band data from the public archive and reduced them.
We also used the WIRCam on CFHT to obtain a deep $K_S$-band image with nearly 50 hr of integration.
Lastly, we obtained a deep $K_S$-band image of the GOODS-N
with the Multi-Object Infrared Camera and Spectrograph (MOIRCS) on the 8.2~m Subaru
Telescope.  This MOIRCS imaging is described in \citet{wang09a}.  
In this imaging, GOODS 850--3 received approximately 4 hr of integration.  
All the above $J$- and $K_S$-band images are still being deepened by various groups.
All the data reduction was carried out by our group using SIMPLE Imaging and Mosaicking Pipeline
(SIMPLE\footnote{also see \url{http://www.asiaa.sinica.edu.tw/$\sim$whwang/idl/SIMPLE}}, \citealp{wang09b}). 
More details about the $K_S$-band observations and data reduction can be found in \citet{barger08} and \citet{wang09b}.
The reduction of the $J$-band data is identical to that of the $K_S$-band.
Here we adopt the MOIRCS $K_S$-band image for morphological analyses because of its high
angular resolution (FWHM $\sim0\farcs4$), and the WIRCam $K_S$-band image for photometry 
because of its greater depth.  

We present the ultradeep ACS white, WIRCam $J$, and MOIRCS $K_S$ images of GOODS~850--3 in 
Figure~3, and the F435W to 8.0 $\mu$m photometry in Table~2.  
The ACS fluxes were measured with $d=0\farcs8$ apertures at the location of the $K_S$ source.  
This is the maximum possible aperture size for the ACS fluxes to be free of contamination from nearby
objects. The $J$-band image does not have sufficient resolution to separate GOODS~850--3 from
nearby objects, and therefore we did not attempt to measure its $J$-band flux.  The $K_S$-band flux
is measured with SExtractor \citep{bertin96} ``AUTO'' aperture, which approximates its 
total flux.  In addition, \citet{wang09b} used a CLEAN-like method to construct \emph{Spitzer} IRAC 
source catalogs based on the WIRCam $K_S$ image. We include their IRAC fluxes for GOODS~850--3
in Table~2. We present the observed optical to IRAC SED in Figure~4.

GOODS 850--3 is extremely faint in the optical, undetected by even the combination of
the four ACS images (also see \citealp{chap04}).
It shows some hint of flux at F850LP (see Table~2) but nothing appears visually in
the F850LP image and the white image.
It clearly shows up at $J$-band, however it is blended with a nearby galaxy.
In the $K_S$-band and \emph{Spitzer} images, GOODS 850--3 is the brightest galaxy in its neighborhood.
With the excellent, $0\farcs4$, resolution of Subaru at $K_S$-band, the stellar component of GOODS~850--3 is resolved.
A simple Gaussian fit to the source in this $K_S$-band image gives a FWHM of $1\farcs03 \times 0\farcs63$,
corresponding to $8.7\times5.3$ kpc$^2$ at its redshift. 
The image shows an elongated morphology, which may be indicative of an edge on rotating disk, and a pronounced
central region that coincides with the radio position, but does not dominate the total $K_S$ luminosity.  
These are consistent with a nuclear starburst hosted by a massive galaxy.

The above optical and near-IR observations imply a large galaxy of roughly 9~kpc
in size that is entirely hidden by dust. This shows that dense molecular clouds are
widely distributed in this galaxy, despite the fact that only its nucleus is actively forming
stars.

\subsection {Star Formation Rate and Surface Density}

The results obtained from various bands in the electromagnetic spectrum (radio to the X-ray) suggest that the
dominant power source in the SMG GOODS~850--3 is a startburst. Therefore, in the following, we derive
an estimate of its star formation rate and star formation surface density.

The 850~$\mu$m
flux density of GOODS~850--3 implies a total IR luminosity of $1.5\times10^{13}L_\sun$
($L_{\rm IR}=1.9\times10^{12}S_{\rm 850 \mu m} L_\sun$ mJy$^{-1}$; \citealp{blain02}).
Assuming that this IR luminosity is entirely powered by a starburst, then the associated star formation
rate (SFR) is $\sim 2500~M_\sun$ yr$^{-1}$ for a Salpeter initial mass function
(${\rm SFR}(M_\sun$ yr$^{-1})=1.7\times10^{-10} L_{\rm IR}(L_{\sun})$; \citealp{kennicutt98}).
An estimate of the SFR can also be calculated using the conversion factor between radio luminosity and SFR
(${\rm SFR}(M_\sun$ yr$^{-1})=5.9 \times10^{-22} L_{\rm 1.4GHz}({\rm W~Hz^{-1}})$; \citealp{yun01}), which is
derived using the local radio--FIR correlation. The radio luminosity of GOODS~850--3 is $3.7 \times 10^{24}$~W~Hz$^{-1}$,
and the resulting SFR is $\sim 2200~M_\sun$ yr$^{-1}$. The consistency in the SFR values
derived from the radio and the IR luminosities simply confirms that this SMG follows the local radio-FIR
correlation. Moreover, this suggests that the star formation activity is well confined within the radio source detected
in our HSA observations.

The size of the source seen in our 1.4 GHz image (Figure~2) compares well with other SMGs, but is considerably smaller
than optically selected starbursting galaxies at high redshift \citep[see, e.g., ][]{bouche07}.
From the size and the SFR of this SMG, we derive a star formation rate surface density of
$\Sigma_{\rm SFR} \sim700~M_\sun$~yr$^{-1}$~kpc$^{-2}$. This surface density is extremely high
even after including the $\sim40\%$ uncertainty in its size. 
For instance, the derived $\Sigma_{\rm SFR}$ in GOODS~850--3
is a few times larger than that seen in a sample of SMGs imaged at millimeter wavelengths \citep{tac06}.
Such a high value is at the very high end of SMGs (see Figure~3 of \citealp{bouche07}), and
is comparable to that measured in the two
extremely luminous SMGs resolved by high resolution submillimeter observations \citep{younger}. Furthermore,
the $\Sigma_{\rm SFR}$ in GOODS~850--3 is comparable to that measured in the host galaxy of the $z=6.42$ QSO
SDSS~J114816.64+525150.3 \citep{walter09}. Such high $\Sigma_{\rm SFR}$ values are
consistent with recent theoretical descriptions of
(dust opacity) Eddington limited star formation of a radiation pressure-supperted starbursts on kpc scales
\citep{tt05}.

Using the population synthesis model of \citet{bruzual03}, the Hyperz package \citep{bolzonella00}
package, and the ACS, $K_S$, and IRAC photometry in Table~2, we find a stellar mass of $1.4\times10^{11}$ $M_\sun$
in GOODS~850--3. The best-fit SED is shown in Figure~4.
With a star formation rate of $\sim2500$ $M_\sun$ yr$^{-1}$, the stellar mass can be
doubled in just 56 Myr, roughly 1.6\% of its Hubble time. The model also requires an
extinction of $A_V\sim2.8$ to explain the nondetections in the optical bands.
We expect the extinction in the nuclear starburst component to be even larger.
Furthermore, we derive a value of 18~Gyr$^{-1}$ for the star-formation rate per unit
stellar mass (i.e., specific star-formation rate; SSFR) in GOODS~850--3.
This is an order of magnitude higher than that seen in normal star forming galaxies
at $z\sim2$, which have SSFR values of about $2-3$~Gyr$^{-1}$ \citep{pan09}.

In summary, GOODS~850--3 has a kpc-scale nuclear starburst with a star formation rate of
$\sim2500$ $M_\sun$ yr$^{-1}$.
No evidence of an AGN is seen at any wavelength. An upper limit on its MIR AGN
contribution of 50\% is estimated by \citet{pope08}, but the AGN contribution to its
total IR luminosity should be much lower. Moreover, our HSA observations
reveal that more than 90\% of the radio continuum emission in this source is extended and
not confined to a central compact AGN.
GOODS~850--3 has a stellar component of $1.4\times10^{11}$ $M_\sun$ with a spatial extent of
roughly 9 kpc. This huge and massive stellar component is entirely hidden by dust,
suggesting an extremely rich molecular gas reservoir fueling the nuclear starburst.

\section{ACKNOWLEDGMENTS}

This research made use of the NASA/IPAC
Extragalactic Database (NED), which is operated by the Jet Propulsion
Laboratory, California Institute of Technology, under contract with the
National Aeronautics and Space Administration.
This work was initiated when WHW was a Jansky Fellow at NRAO. WHW
acknowledges the great support from the NRAO, and LLC and AJB acknowledge
support from NSF grants AST 0709356 and AST 0708793, respectively.

\begin{figure}
\epsscale{1}
\plotone{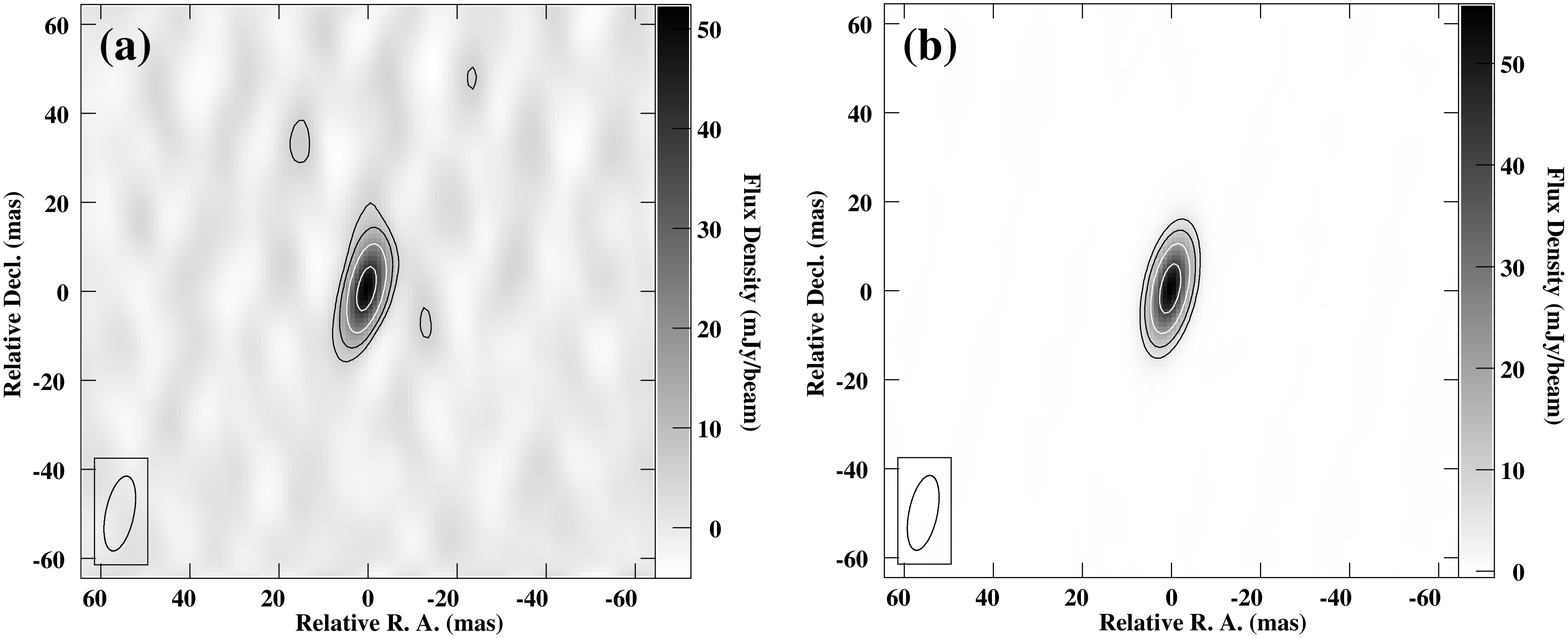}
\figcaption{
Continuum images of the phase-check calibrator J1219+6600 at 1.4~GHz: a)
obtained by applying the phase and the amplitude self-calibration solutions of the phase
reference source J1229+6335, b) obtained by self calibrating J1219+6600 itself, in both
phase and amplitude. The restoring beam size in both images is $17.1 \times 6.4$~mas in position
angle $-12{\degr}$. The contour levels are at $-$3, 3, 6, 12, 24 times the rms noise level in the
phase-referenced image (left), which is 1.71~mJy~beam$^{-1}$. The gray-scale range is indicated
at the right side of each image. The reference point (0, 0) in both images
is $\alpha(\rm{J2000.0})= 12^{\rm h}19^{\rm m}35\rlap{.}^{\rm s}7941$,
$\delta(\rm{J2000.0})=+66^{\circ}00^{'}31\rlap{.}^{''}845$. 
\label{f1}}
\end{figure}

\clearpage
\begin{figure}
\epsscale{1}
\plotone{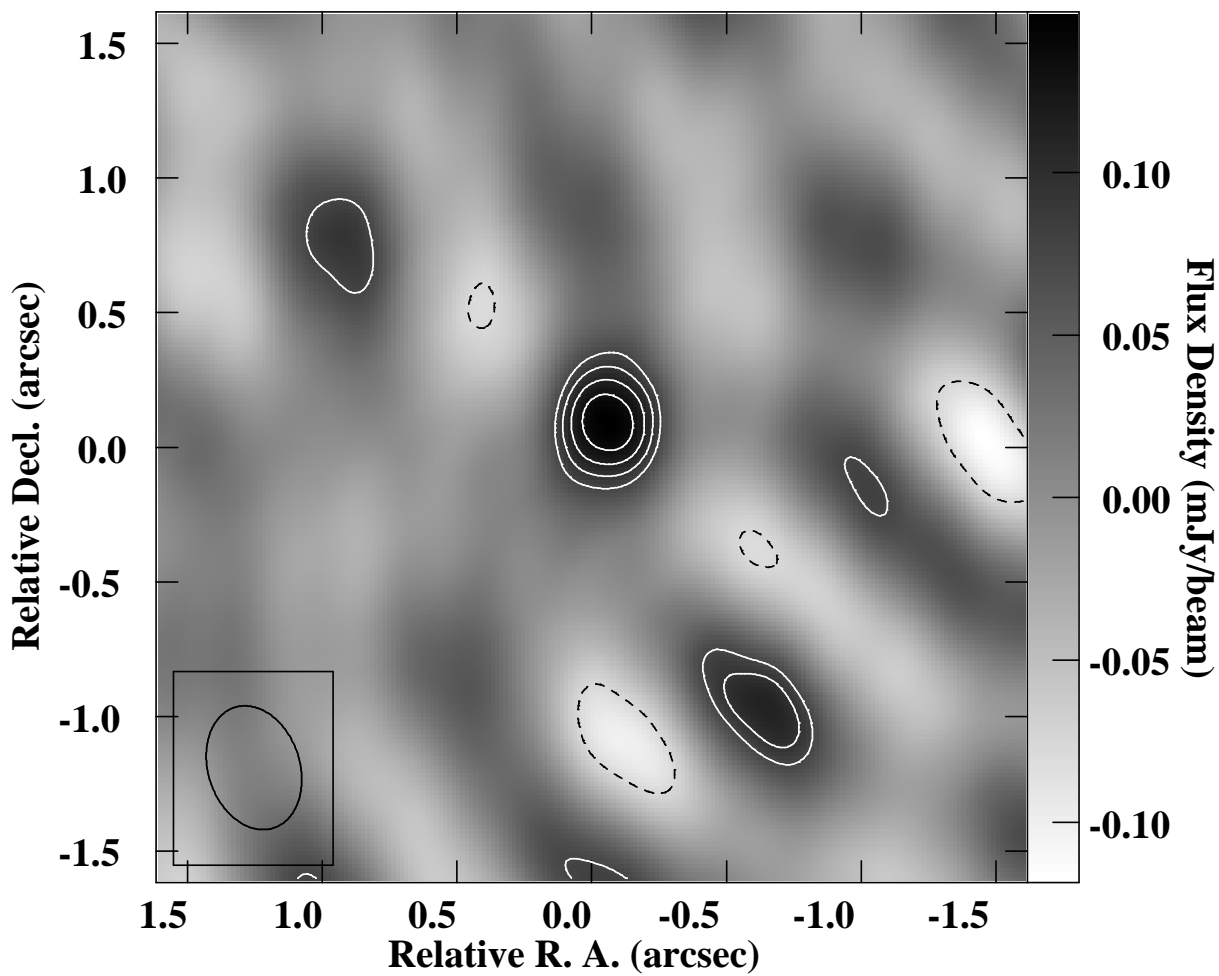}
\figcaption{
Naturally weighted 1.4~GHz continuum image of GOODS~850--3 at $0.47'' \times 0.34''$
resolution (P.A.=$18\degr$). The peak flux density is 148~$\mu$Jy~beam$^{-1}$, and the contour levels are
at $-$2, 2, 2.5, 3, 3.5 times the rms noise level, which is 38~$\mu$Jy~beam$^{-1}$. The gray-scale range
is indicated at the right side of the image. The reference point (0, 0)
is $\alpha(\rm{J2000.0})= 12^{\rm h}36^{\rm m}18\rlap{.}^{\rm s}3317$,
$\delta(\rm{J2000.0})=+62^{\circ}15^{'}50\rlap{.}^{''}608$. 
A two dimensional Gaussian taper falling to 30\% at 0.5~M$\lambda$ was applied.
\label{f2}}
\end{figure}

\clearpage
\begin{figure}
\epsscale{1}
\plotone{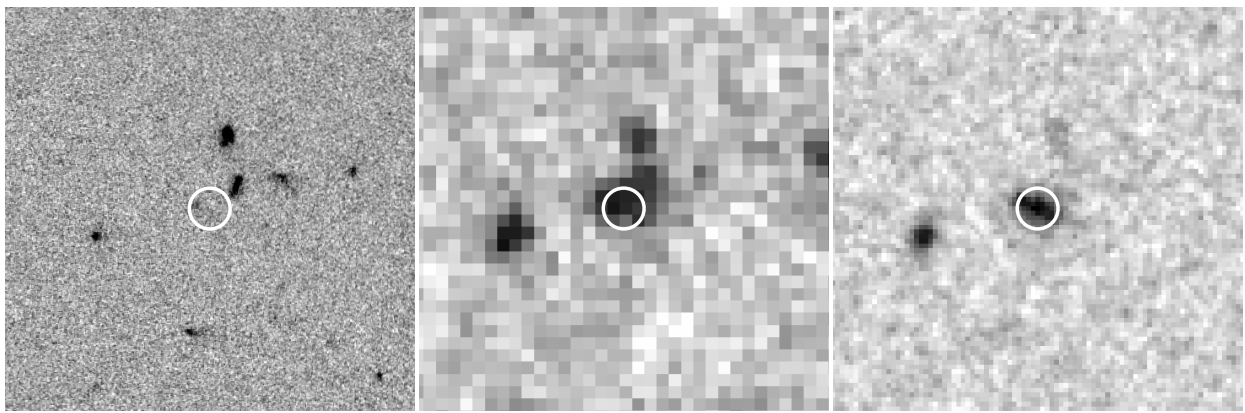}
\figcaption{Optical and near-IR images of GOODS~850--3. From left are the \emph{HST} ACS
``white'' image (F435W+F606W+F775W+F850LP), the CFHT $J$-band image, and the Subaru $K_S$-band image. All panels
have $10\arcsec$ sizes and north is up. Circles indicate the 1.4 GHz position
and have $1\arcsec$ diameters.  Note that GOODS~850--3 entirely disappears in the
optical but becomes very bright at longer wavelengths.
\label{f3}}
\end{figure}

\clearpage
\begin{figure}
\epsscale{1}
\plotone{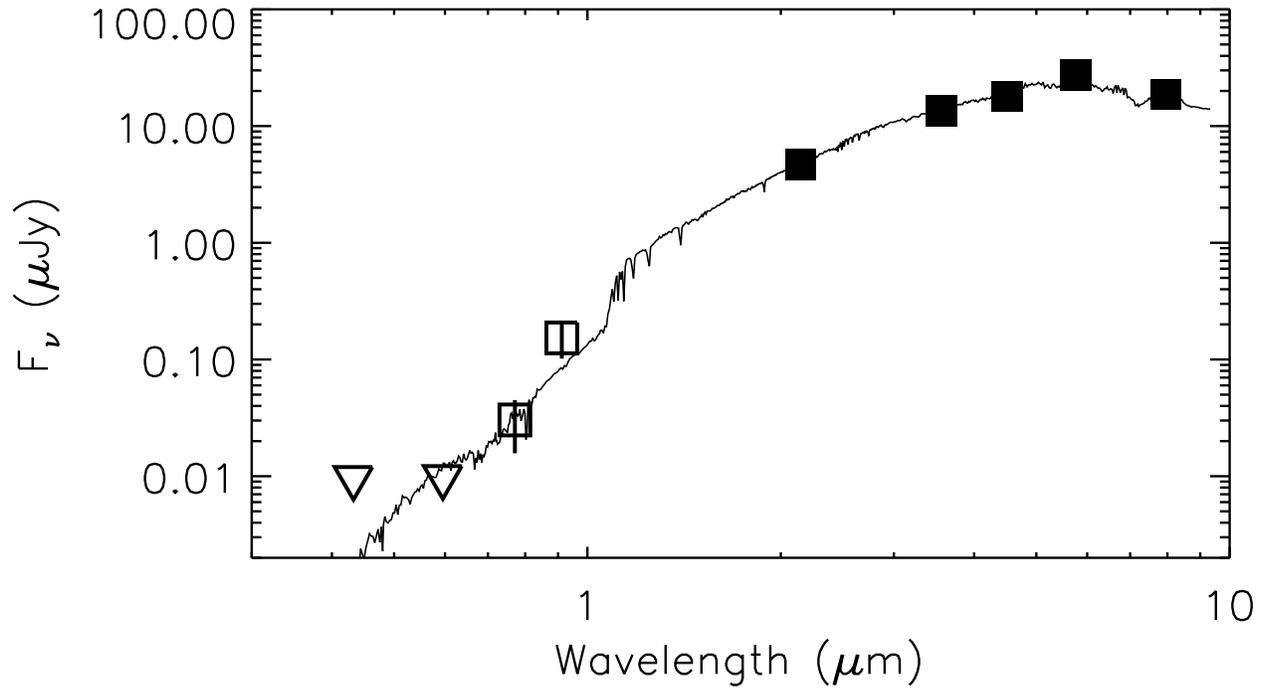}
\figcaption{The SED of GOODS 850--3.  Symbols show the observed data in Table~2.
Filled squares are detections.  Open squares are low significance flux measurements.
Downward triangles are 1 $\sigma$ upper limits.  The curve is the best-fit
SED from the \citet{bruzual03} models, which has a continuous starburst of
0.36 Gyr, extinction of $A_V=2.8$, and a stellar mass of $1.4\times10^{11}M_\sun$.
\label{f4}}
\end{figure}

\clearpage
\oddsidemargin=-1cm

\begin{deluxetable}{lc}
\tablenum{1}
\tablecolumns{6}\tablewidth{0pc}
\tablecaption{P{\footnotesize ARAMETERS} {\footnotesize OF THE} VLBI
O{\footnotesize BSERVATIONS} {\footnotesize OF} GOODS~850--3}
\tablehead{\colhead{Parameters} & \colhead{Values}}
\startdata
Observing Dates \dotfill  & 2006 Dec. 9 \& 2007 Feb. 13 \\
Total observing time (hr)\dotfill  & 16 \\
Phase calibrator\dotfill  & J1229+6335 \\
Phase-referencing cycle time (sec)\dotfill  &  280 \\
Frequency (GHz)\dotfill  &  1.4 \\
Total bandwidth (MHz)\dotfill   & 64\\
\enddata
\end{deluxetable}

\begin{deluxetable}{lc}
\tablecolumns{2}\tablenum{2}
\tablecaption{O{\footnotesize PTICAL TO}  M{\footnotesize ID-}I{\footnotesize NFRARED} 
P{\footnotesize HOTOMETRIC} D{\footnotesize ATA OF} GOODS~850--3}
\tablewidth{0pt}
\tablehead{\colhead{Wave Band} & \colhead{Flux ($\mu$Jy)} }
\startdata
ACS F435W \dotfill & 		$0.012\pm 0.010$  \\
ACS F606W \dotfill & 		$0.008 \pm 0.010$  \\
ACS F775W \dotfill & 		$0.030 \pm 0.015$  \\
ACS F850LP \dotfill &		$0.153 \pm 0.051$  \\
WIRCam $K_s$ \dotfill &    $4.71\pm0.14$ \\
IRAC 3.6 $\mu$m \dotfill &	$13.4 \pm 0.2$  \\
IRAC 4.5 $\mu$m \dotfill &	$17.9 \pm 0.2$  \\
IRAC 5.8 $\mu$m \dotfill &	$27.3 \pm 0.6$  \\
IRAC 8.0 $\mu$m \dotfill &	$18.5 \pm 0.6$ 
\enddata
\end{deluxetable}

\end{document}